\documentclass[12pt]{iopart}

\usepackage{bm}
\usepackage{graphicx}
\begin{document}

\title{Quantum theory of multimode fields: Applications to optical resonators}

\author{C Viviescas and G Hackenbroich}
\address{Universit\"at Duisburg--Essen, Fachbereich Physik, 45117
  Essen, Germany}

\begin{abstract}
  A recently developed technique for the system--and--bath
  quantization of open optical cavities is applied to three resonator
  geometries: A one dimensional dielectric, a Fabry--Perot resonator,
  and a dielectric disk. The system--and--bath Hamiltonian for these
  geometries is derived starting from Maxwell's equations and employed
  to compute the electromagnetic fields, the resonances, and the
  cavity gain factors. Exact agreement is found with standard
  quantization methods based on a modes--of--the--universe
  description. Our analysis provides a microscopic justification for
  the system--and--bath quantization even in the regime of spectrally
  overlapping modes. Combined with random--matrix theory our
  quantization method can serve as a starting point for a quantum
  theory of wave--chaotic and disordered optical media.
\end{abstract}


\section{Introduction}
A standard description of losses in quantum mechanical systems is the
system--and--bath model: One considers the system of interest coupled
to a large quantum system acting as a reservoir or bath.  The
reservoir is assumed to be so large that its state is not much
affected by the coupling to the system. In contrast, the coupling to
the reservoir introduces both losses and noise for the system
dynamics. In optical resonators the system--and--bath model has been
used for more than 40 years \cite{Seni59,Seni60,Gard00} as a
phenomenological description for the leakage of radiation into the
external electromagnetic field.

In two recent papers \cite{Hack02,Vivi03} we reported a
\emph{microscopic} derivation of the system--and--bath model from
Maxwell's equations and derived explicit expressions for the coupling
amplitudes between system and bath modes. We demonstrated that the
model not only applies to good resonators with spectrally
well--separated modes but that it also provides an \emph{exact}
description for spectrally--overlapping or even overdamped modes. The
field quantization in the presence of mode overlap has generated a
substantial literature in recent years \cite{Grun96,Dalt99,Dutr00b}.
The motivation for our work came from recent studies of unstable
optical cavities \cite{Hame89,Eijk96,Lind98} and from experiments on
strongly disordered amplifying media \cite{Frol99,Cao99,Cao01,Cao03},
so-called random lasers. The losses in such lasers are typically much
larger than in traditional lasers as the light is not confined by
mirrors but by the multiple chaotic scattering within the disordered
medium. The amount of disorder determines both the laser mode
amplitudes and frequencies. Therefore, the modes of random media
depend on the statistical properties of the underlying random medium,
and must be analyzed in a statistical fashion. Traditional methods of
field--quantization, such as the so--called modes--of--the--universe
approach \cite{Lang73}, are not suited for a statistical description
as they do not provide explicit information about the field inside the
resonator region. In contrast, statistics naturally enters our
system--and--bath model \cite{Hack02,Vivi03} when the internal system
dynamics is modeled by random--matrix theory \cite{Guhr98}.  In our
previous work we have focused on the general derivation of the
system--and--bath model, the discussion of the resulting system
dynamics, and the connection with random--matrix theory. It is the
purpose of the present paper to demonstrate our method explicitly for
a number of models frequently used for optical resonators.

\begin{figure}[t]
  \begin{center} \includegraphics[scale=1]{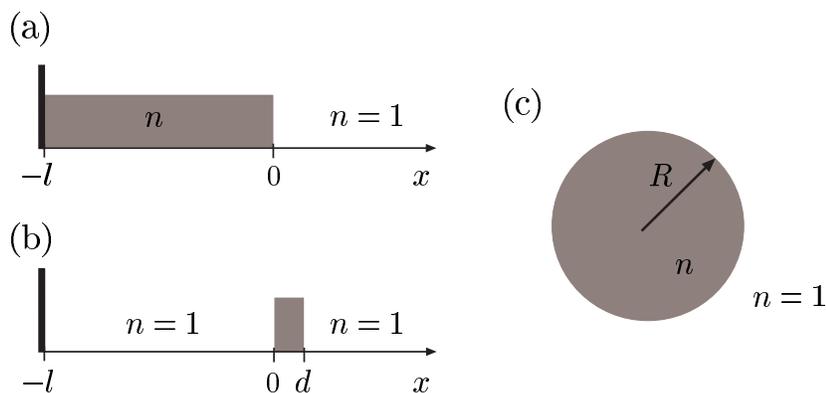}
    \caption{\label{fig:cavities} Open optical cavities studied 
      in this paper: (a) A one--dimensional dielectric slab of
      length $l$ with (positive) refractive index $n$ bounded on
      one side $(x=-l)$ by a perfectly reflecting mirror. (b) A
      one--dimensional cavity bounded by a perfectly reflecting
      mirror at $(x=-l)$ and a thin semi--transparent mirror of
      width $d<<l$ at $(x=0)$. (c) A two--dimensional dielectric
      disk with refractive index $n$.}
\end{center}
\end{figure}

Specifically, we consider the three types of cavities shown in
\fref{fig:cavities}: (a) A one--dimensional dielectric slab
with (positive) refractive index $n$ bounded on one side by a
perfectly reflecting mirror, (b) a one--dimensional cavity defined
by a perfectly reflecting mirror on one side and a thin
semi--transparent mirror on the other side, and (c) a
two--dimensional dielectric disk with refractive index $n$. The
dielectrics are embedded in empty space. The normal modes of all
three systems in question as well as their scattering properties
can be computed exactly. Likewise, exact expressions can be
obtained for the electromagnetic fields both within the cavities
and in the external region. This makes all three models an ideal
testing ground for our system--and--bath description. According to
that description the Hamiltonian of the complete system, comprising
the resonator and the external region, can be represented as
\begin{eqnarray}
\label{eq:hamfesh}
H &= \sum_{\lambda} \hbar \omega_{\lambda} a^{\dagger}_{\lambda}
a_{\lambda} + \sum_m \int\limits_{\omega_{m}}^{\infty} \! \rmd\omega \,
\hbar\omega b_m^{\dagger} (\omega) b_m(\omega) \nonumber\\
&\quad + \hbar
\sum_{\lambda} \sum_m \int\limits_{\omega_{m}}^{\infty} \!  \rmd\omega
\left[ \mathcal{W}_{\lambda m}(\omega) \, a^{\dagger}_{\lambda}
  b_{m}(\omega) +\mathcal{V}_{\lambda m}(\omega) \, a_{\lambda}
  b_{m}(\omega) + {\rm H.c. } \right].
\end{eqnarray}
There is a discrete set of cavity modes with index $\lambda$, and a
continuous set of external modes labeled by the frequency $\omega$ and
the discrete index $m$, which specifies the asymptotic boundary
conditions far from the resonator (including polarization);
$\omega_{m}$ is the frequency threshold above which external modes
with the ``channel''--index $m$ exist. The operators $a_\lambda$ and
$a_{\lambda}^{\dagger}$ are, respectively, the bosonic annihilation
and creation operators associated with the cavity modes. They satisfy
the standard commutation relations $[a_\lambda,a_{\lambda '}]=0$,
$[a_\lambda,a_{\lambda '}^{\dagger}]= \delta_{\lambda
\lambda'}$. Likewise, the channel operators $b_m(\omega)$ and
$b_m^{\dagger}(\omega)$ satisfy the commutation relations
$[b_m(\omega),b_n(\omega')]=0$, $[b_m(\omega), b_n^{\dagger}(\omega')]
=\delta_{mn}\,\delta(\omega-\omega')$.  Operators related to different
subsystems commute. Explicit expressions for the coupling amplitudes
$\mathcal{W}_{\lambda m}(\omega)$, $\mathcal{V}_{\lambda m}(\omega)$
will be given below.

The electric and magnetic fields are exactly represented within the
system--and--bath approach by
\numparts
\begin{eqnarray}
\label{eq:EFf}
{\bm E} ({\bm r},t) &= \rmi \sum_\lambda \left(\frac{\hbar
  \omega_\lambda}{2} \right)^{1/2} \bigl[ a_\lambda(t) {\bm
    u}_\lambda ({\bm r}) - {\rm H.c.} \bigr] \nonumber\\
&\quad + \rmi \sum_{m}
\int\limits_{\omega_{m}}^{\infty} \! \rmd\omega
\left(\frac{\hbar\omega}{2} \right)^{1/2} \bigl[
  b_{m}(\omega,t) {\bm v}_{m}(\omega,{\bm r}) - {\rm H.c.} \bigr] \,
, \\
\label{eq:MFf}
{\bm B} ({\bm r},t) &= c \sum_\lambda \left(\frac{\hbar
}{2\omega_\lambda} \right)^{1/2} \biggl[ a_\lambda(t)
  \bigl(\nabla \times {\bm u}_\lambda \bigr) ({\bm r}) + {\rm H.c.}
  \biggr] \nonumber \\ 
&\quad + c \sum_{m}
\int\limits_{\omega_{m}}^{\infty} \! \rmd\omega
\left(\frac{\hbar}{2\omega} \right)^{1/2} \biggl[
  b_{m}(\omega,t) \bigl(\nabla \times{\bm v}_{m}(\omega,{\bm r})\bigr)
  + {\rm H.c.} \biggr],
\end{eqnarray}
\endnumparts where $c$ is the velocity of light in vacuum. The modes
${\bm u}_\lambda ({\bm r})$ and ${\bm v}_{m} (\omega,{\bm r})$ have
support only in the resonator and the external region, respectively.
Although the expansions (\ref{eq:EFf}) and (\ref{eq:MFf}) reduce
precisely to the standard expressions known from closed systems, the
field \emph{dynamics} given by Hamiltonian (\ref{eq:hamfesh}) is
fundamentally different due to the coupling between the resonator and
the external region \cite{Vivi03,Hack03}.

The microscopic formulation of the system--and--bath model must
include a definition of the system modes ${\bm u}_\lambda$ and the
bath modes ${\bm v}_m (\omega)$. We will specify proper definitions
below for the three systems of interest, but note at this point that
these definitions (and the definition of the system--and--bath
Hamiltonian itself) entail some degree of arbitrariness. This
arbitrariness arises generically for open resonators; its origin is
that the separation of space into two regions ``inside'' and
``outside'' the resonator is not unique. Consider for example a cavity
with an opening, i.e.\ a hole in the material walls of the cavity
boundary. Any choice of a fictitious surface covering the hole yields
its own inside/outside separation. Moreover, different boundary
conditions may be imposed at the chosen separating surface.
Nevertheless, each such surface and boundary condition entail
eigenmodes allowing to represent the electromagnetic field almost
everywhere. We emphasize that the expansion cannot be expected to
converge uniformly or even pointwise, and in particular not on an
arbitrarily chosen boundary. Still the system-and--bath representation
is exact in the $L^2$ sense required by quantum mechanics. The freedom
in the choice of the separating surface and the boundary conditions
manifests itself in the Hamiltonian (\ref{eq:hamfesh}) and the field
expansions (\ref{eq:EFf}) and (\ref{eq:MFf}): the internal frequencies
$\omega_{\lambda}$, the coupling amplitudes $\mathcal{W}$ and
$\mathcal{V}$, and the mode functions ${\bm u}$ and ${\bm v}$, all
depend explicitly on these choices.  However, we show below that all
physical observables, in particular the electromagnetic fields and the
scattering amplitudes, do not rely on these choices and only depend on
the physical boundary conditions imposed by Maxwell's equations.

The outline of the paper is as follows. In \sref{sec:basics} we
collect the main results of our system--and--bath approach relevant
for resonant cavities. In sections~\ref{sec:ex01}--\ref{sec:ex04} the
method is applied to the three resonators introduced earlier. We
compute the respective system--and--bath Hamiltonians, and show that
the electromagnetic fields and the scattering properties agree exactly
with results obtained by direct solutions of the quantization problem.
Performing the computations for different sets of boundary conditions
along the surface separating the resonator and channel region, we
demonstrate that the physical observables are independent of that
choice of boundary condition. We conclude in \sref{sec:conclusions}.

\section{System--and--bath quantization}
\label{sec:basics}

In two recent papers \cite{Hack02,Vivi03} we developed a
microscopic system--and--bath approach to the field quantization in
open optical cavities. To keep the present paper self--contained we
summarize the relevant results in the present section.  Our
quantization technique applies to linear dielectric media
characterized by a scalar dielectric constant $\epsilon({\bm r})$
that may explicitly depend on position and that we assume real and
frequency independent.  Working in the Coulomb gauge in the absence
of charges, the exact eigenmodes ${\bm f}_m(\omega,{\bm r})$ of
Maxwell's equations \cite{Glau91} are solutions of the Helmholtz
equation
\begin{equation}
\label{eq:evequniv}
\nabla \times \left[ \nabla \times {\bm f}_m(\omega,{\bm r})
  \right] - \frac{\epsilon({\bm r})\omega^2}{c^2} {\bm
  f}_m(\omega,{\bm r})={\bm 0} ,
\end{equation}
and satisfy the generalized transversality condition $ \nabla \cdot
\left[\epsilon({\bm r}) {\bm f}_m(\omega, {\bm r}) \right] = 0$.  The
modes are labeled by the continuous frequency $\omega$ and a discrete
index $m$; the latter specifies the asymptotic conditions far away
from the resonator. For example, these conditions could correspond to
a scattering problem with incoming and outgoing waves. Then ${\bm
  f}_m(\omega,{\bm r})$ represents a solution with an incoming wave
only in channel $m$ and outgoing waves in all scattering
channels. Using notation from scattering theory, the region outside
the resonator will be called channel region below.  We follow
reference \cite{Glau91} and stick to the usual internal product with
respect to which the differential operator in equation
\eref{eq:evequniv} is not Hermitian. However, the substitution
\begin{equation}
\label{eq:fguniv}
{\bm f}_{m}(\omega,{\bm r}) = \frac{1}{\sqrt{\epsilon({\bm r})}}
\bm{\phi}_{m} (\omega,{\bm r})
\end{equation}
transforms equation \eref{eq:evequniv} into the hermitian eigenvalue
problem
\begin{equation}
\label{eq:hevequniv}
L \bm{\phi}_{m} (\omega, {\bm r}) \equiv
\frac{1} {\sqrt{\epsilon({\bm r})}} \nabla \times 
\left( \nabla \times \frac{ \bm{\phi}_{m} (\omega, {\bm r})}
{\sqrt{\epsilon({\bm r})}} \right) =
\frac{\omega^2} {c^2} \bm{\phi}_{m} (\omega, {\bm r}) ,
\end{equation}
for the operator $L$. The functions $\bm{\phi}_{m} (\omega)$ provide a
complete orthonormal set in the space of $L^2$ functions. It follows
from equation \eref{eq:fguniv} that the functions ${\bm f}_m(\omega)$
satisfy the orthonormalization condition
\begin{equation}
\label{eq:normuniv}
\int \! \rmd{\bm r} \,\epsilon({\bm r}) {\bm f}_m(\omega, {\bm r}) {\bm
f}_{n}(\omega', {\bm r}) = \delta_{mn}\,\delta(\omega - \omega').
\end{equation}
The quantization of the electromagnetic field in terms of
eigenmodes of the Helmholtz equation is known as the
modes--of--the--universe approach \cite{Lang73,Knol91,Glau91}.
With\-in this approach the field Hamiltonian reduces to a sum of
independent harmonic oscillators
\begin{equation}
\label{eq:hamuniv}
H = \frac{1}{2} \sum_m \int\limits_{\omega_{m}}^{\infty} \! \rmd\omega \,
  \hbar\omega \, \Bigl( A_m^\dagger(\omega) A_m(\omega) +
A_m(\omega) A_m^\dagger(\omega) \Bigr) ,
\end{equation}
and the electromagnetic fields take the standard form
\numparts
\label{eq:fieldsuniv}
\begin{eqnarray}
{\bm E} = \rmi \sum_m \int\limits_{\omega_{m}}^{\infty} \! \rmd\omega \;
  \left( \frac{\hbar\omega}{2}\right)^{1/2} \left[
  A_m(\omega) \rme^{-\rmi\omega t} {\bm f}_m(\omega,{\bm r}) - {\rm H.c.}
  \right],\\ 
{\bm B} = c \sum_m \int\limits_{\omega_{m}}^{\infty} \!
  \rmd\omega \left( \frac{\hbar}{2\omega}\right)^{1/2} \left[
  A_m(\omega) \rme^{-\rmi\omega t} \bigl( \nabla \times {\bm f}_m(\omega,{\bf
  r}) \bigr) + {\rm H.c.} \right].
\end{eqnarray}
\endnumparts
The fundamental disadvantage of the modes--of--the--universe
approach is that it makes no distinction between the cavity and the
external region. As a consequence, the method fails to single out
explicit information about the field inside the cavity.

The system--and--bath technique is based on the separation of the
electromagnetic fields into an inside and an outside contribution.
Likewise, the eigenmodes of the Helmholtz equation are decomposed
into contributions ``living'' inside the resonator or in the
channel space. At a formal level one can achieve the inside/outside
separation by introducing the projection operators
\cite{Fesh62,Vivi03}
\begin{equation}
\label{eq:proj}
\mathcal{Q} = \int_{{\bm r} \in I} \! \rmd{\bm r}\, |{\bm r} 
\rangle\langle{\bm r}| \quad {\rm and} \quad
\mathcal{P} = \int_{{\bm r} \not\in I} \! \rmd{\bm r}\, |{\bm r}
\rangle\langle{\bm r}|,
\end{equation}
where $|{\bm r} \rangle$ denotes a standard position eigenket and
$I$ represents the resonator region. The projection operators are
orthogonal, $\mathcal{QP} = \mathcal{PQ} = 0$, and complete,
$\mathcal{Q} + \mathcal{P} = 1$. Thus an arbitrary function ${\bm
  \phi}$ in Hilbert space may be decomposed into its projections
onto the resonator and channel space
\begin{equation}
\label{eq:split}
|{\bm \phi}\rangle = \mathcal{Q}|{\bm \phi}\rangle + 
\mathcal{P}|{\bm \phi}\rangle \equiv |{\bm \mu}\rangle + 
|{\bm \nu}\rangle.
\end{equation}
Acting on $|{\bm \phi}\rangle$ with the operator $L$ we obtain
\begin{equation}
\label{eq:Lpro}
L|{\bm \phi}\rangle = L_{\mathcal{QQ}} |{\bm \mu}\rangle +
L_{\mathcal{QP}} |{\bm \nu}\rangle + L_{\mathcal{PQ}} |{\bm \mu}\rangle +
L_{\mathcal{PP}} |{\bm \nu}\rangle ,
\end{equation}
where $L_{\mathcal{QQ}}$ and $L_{\mathcal{PP}}$ are the projections
of $L$ onto the resonator and channel space, and $L_{\mathcal{QP}}$
and $L_{\mathcal{PQ}}$ the coupling terms. As explained in the
introduction and illustrated in references \cite{Vivi03,Savi03}, this
decomposition is by no means unique: different choices of the
separating surface and different boundary conditions along that
surface result in different decompositions of $L$. However, all
those decompositions are subject to the condition that $L$ remains
Hermitian, i.e.\ $L_{\mathcal{QQ}}= L_{\mathcal{QQ}}^{\dagger}$,
$L_{\mathcal{PP}}= L_{\mathcal{PP}}^{\dagger}$ and
$L_{\mathcal{QP}}= L_{\mathcal{PQ}}^{\dagger}$.

The projections $L_{\mathcal{QQ}}$ and $L_{\mathcal{PP}}$ define
Hermitian eigenvalue problems for the isolated cavity and channel
region, respectively. The closed resonator modes $|{\bm
  \mu}_{\lambda}\rangle$ are the solutions of the eigenvalue
problem
\begin{equation}
L_{\mathcal{QQ}} |{\bm \mu}_{\lambda}\rangle =
\left(\frac{\omega_{\lambda}}{c} \right)^2 |{\bm \mu}_{\lambda}\rangle,
\end{equation}
and form a discrete orthonormal basis for the cavity subspace. 
Likewise, the channels modes $|{\bm \nu}_{m}(\omega)\rangle$
satisfy the equation
\begin{equation}
\label{eq:nug}
L_{\mathcal{PP}} |{\bm
  \nu}_{m}(\omega)\rangle = \left(\frac{\omega}{c} \right)^2 |{\bm
  \nu}_{m}(\omega)\rangle,
\end{equation}
and constitute a continuous basis for the channel region. We note that
the functions ${\bm \mu}_{\lambda}({\bm r})$ have support only inside
the resonator, while the functions ${\bm \nu}_{m}(\omega,{\bm r})$ are
nonzero only in the channel region.

It follows from these definitions that the resonator and channel
modes form a complete set of Hilbert space functions in the
respective subregions. One may therefore use these functions to
expand the eigenstates of the Helmholtz equation
\numparts
\begin{eqnarray}
\label{eq:phisplit}
{\bm \phi}_{m} (\omega,{\bm r}) & = \sum_{\lambda} \alpha_{m\lambda}
(\omega) {\bm \mu}_{\lambda}({\bm r}) + \sum_{m'}
\int\limits_{\omega_{m'}}^{\infty} \! \rmd\omega^\prime \; \beta_{mm'}
(\omega,\omega^\prime) {\bm \nu}_{m'} (\omega^\prime,{\bm r}), \\
\label{eq:fsplit}
{\bm f}_{m} (\omega,{\bm r}) & = \sum_{\lambda} \alpha_{m\lambda}
(\omega) {\bm u}_{\lambda}({\bm r}) + \sum_{m'}
\int\limits_{\omega_{m'}}^{\infty} \! \rmd\omega^\prime \; \beta_{mm'}
(\omega,\omega^\prime) {\bm v}_{m'} (\omega^\prime,{\bm r}),
\end{eqnarray}
\endnumparts
where we introduced the functions ${\bm u}_{\lambda}({\bm r})
\equiv {\bm \mu}_{\lambda} ({\bm r}) / \sqrt{\epsilon({\bm r})}$
and $ {\bm v}_{m} (\omega, {\bm r}) \equiv {\bm \nu}_{m}
(\omega,{\bm r}) / \sqrt{\epsilon({\bm r})}$.  The expansion
coefficients can be recovered from the mode functions using the
relations
\numparts
\begin{eqnarray}
\label{eq:expcoefa}
\alpha_{m\lambda}(\omega) = \langle {\bm \mu}_{\lambda}| {\bm
\phi}_{m} (\omega) \rangle, \\
\label{eq:expcoefb}
\beta_{mm'} (\omega,\omega') = \langle {\bm \nu}_{m'}
(\omega^\prime)|{\bm \phi}_{m} (\omega)\rangle.
\end{eqnarray}
\endnumparts

The system--and--bath Hamiltonian (\ref{eq:hamfesh}) and the field
representations (\ref{eq:EFf}) and (\ref{eq:MFf}) follow upon the
introduction of bosonic creation and annihilation operators $a$ and
$a^{\dagger}$ for the cavity modes, and similar operators $b$ and
$b^{\dagger}$ for the channel modes. Finally, the coupling amplitudes
in the Hamiltonian (\ref{eq:hamfesh}) are of the form \numparts
\begin{eqnarray}
\label{eq:coupw}
\mathcal{W}_{\lambda m}(\omega) = \frac{c^2}{2 \sqrt{\omega_\lambda
    \omega}} \langle {\bm \mu}_{\lambda} | L_{\mathcal{QP}} |{\bm
  \nu}_m(\omega) \rangle , \\
\label{eq:coupv}
\mathcal{V}_{\lambda m}(\omega) = \frac{c^2}{2 \sqrt{\omega_\lambda
    \omega}} \langle {\bm \mu}_{\lambda}^{*} | L_{\mathcal{QP}} |{\bm
  \nu}_m(\omega)\rangle .
\end{eqnarray}
\endnumparts
The notation $|{\bm \mu}^{*}\rangle$ means $\langle {\bm r}|{\bm
  \mu}^{*}\rangle \equiv {\bm \mu}^{*}({\bm r})$. For time reversal
invariant systems the wave functions may be chosen real, then the
amplitudes $\mathcal{W}$ and $\mathcal{V}$ become real and
identical, $\mathcal{W} =\mathcal{V}$.

In order to test the system--and--bath approach for the resonators of
interest we compute several physical observables below. Examples
include the electromagnetic fields, and the eigenmodes of the
Helmholtz equation. We also study the cavity resonances, i.e.\ the
complex frequencies that determine the cavity response to external
excitations in the presence of the coupling to the outside world.
Formally the resonances are found as the poles of the resolvent
operator $\mathcal{G}(\omega)$ projected onto the cavity space and
analytically continued in the second Riemann sheet \cite{Cohe92}.
After projection the resolvent can be written as
\begin{equation}
\label{eq:res}
\mathcal{G}_{QQ}(\omega) = \frac{1}{\left(\frac{\omega}{c}\right)^2 -
  L_{\rm eff}(\omega)} \, ,
\end{equation}
where the non-Hermitian operator $L_{\rm eff}(\omega)$ is
expressed through the projections of the differential operator $L$,
\begin{eqnarray}
\label{eq:Leff}
L_{\rm eff}(\omega) &\equiv L_{\mathcal{QQ}} + L_{\mathcal{QP}}
\frac{1}{\left(\frac{\omega}{c}\right)^2 - L_{\mathcal{PP}} + \rmi
  \epsilon}L_{\mathcal{PQ}} \, , \nonumber\\ 
& = L_{\mathcal{QQ}} +
c^2 L_{\mathcal{QP}} \sum_{m} \int\limits_{\omega_{m}}^{\infty} \!
\rmd\omega' \, \frac{|{\bm \nu}_{m}(\omega')\rangle \langle {\bm
    \nu}_{m}(\omega')|}{\omega^2 - \omega'^2 + \rmi
  \epsilon}L_{\mathcal{PQ}},
\end{eqnarray}
here the limit $\epsilon \to 0^{+}$ is implied. The second line
follows upon using the completeness of the channels modes with the
help of equation \eref{eq:nug}. The operator $L_{\rm eff}(\omega)$ is
closely related to the effective Hamiltonian in open quantum systems
\cite{Savi03}. To determine the system resonances one must solve the
eigenvalue problem,
\begin{equation}
\label{eq:Leffeveq}
L_{\rm eff}(\omega) |\xi_{i}(\omega)\rangle = \sigma^2_{i}(\omega)
|\xi_{i}(\omega)\rangle .
\end{equation}
Since $L_{\rm eff}(\omega)$ depends parametrically on $\omega$,
both its right eigenstates $|\xi_{i}(\omega)\rangle$ and the complex
eigenvalues $\sigma_{i}(\omega)$ generally depend on $\omega$ as well.
The states $|\xi_{i}(\omega)\rangle$ correspond to the Kapur--Peierls
states \cite{Kapu38,Kuku89} of scattering theory.  Combining equations
(\ref{eq:res}) and (\ref{eq:Leffeveq}) the cavity resonances are found
as the solutions of the fixed point equation
$c\sigma_{i}(\omega)=\omega$. We show below that this resonance
condition is independent of the inside/outside separation and the
choice of boundary condition made within the system--and--bath
description.

A well--known quantity that shows resonant behavior is the cavity
gain factor
\begin{equation}
\label{eq:cgf}
G^c(\omega) = \frac{\int_{{\bm r}\in I} \! \rmd{\bm r}
  \, \rho(\omega,{\bm r})}{\int_{{\bm r}\in I}\! \rmd{\bm r} \,
  \rho_{0}(\omega,{\bm r})} \, ,
\end{equation}
that is closely related to the dwell time of scattered radiation
inside the resonator \cite{Carv02}. Here, $\rho(\omega, {\bm r})$
is the local density of states inside the cavity and $\rho_0
(\omega,{\bm r})$ the free space local density of states in the
absence of the cavity. The cavity gain factor thus measures the
change in the local density of states introduced by the cavity. For
later use we note that the integrated density of states can be
expressed in terms of the expansions coefficients
(\ref{eq:expcoefa}),
\begin{equation}
\label{eq:aldos}
\int_{{\bm r}\in I} \! \rmd{\bm r} \, \rho(\omega,{\bm r})
= \sum_{m}\sum_{\lambda} |\alpha_{m\lambda}(\omega)|^2.
\end{equation}

\section{One dimensional dielectric cavity}
\label{sec:ex01}

Our first example is the one dimensional dielectric cavity depicted
in \fref{fig:cavities}(a) \cite{Ujih75}. The dielectric with
refractive index $n$ is nonabsorbing and nondispersive. It is
bounded by a perfectly reflecting mirror at $x=-l$ while there is
no mirror at the other end of the dielectric at $x=0$. The free
space outside the cavity runs from $x=0$ to infinity, and light
propagates freely there, $n=1$. We assume the electromagnetic field
to be linearly polarized with the electric field vector pointing in
the $z$--direction. Cavity field excitations will decay due to
leakage into the empty half--space. The dielectric function of the
total system including the cavity and the attached half--space is
given by
\begin{equation}
\label{eq:eps01}
\epsilon(x) = n^2 \Theta(-x)  + \Theta(x),
\end{equation}
where the Heavyside--function $\Theta(x)$ is equal to one for positive
$x$ and vanishes for negative $x$. The exact eigenmodes of Maxwell's
equations for this problem are given in equation \eref{eq:f01}. To solve the
problem within the system--and--bath approach, we separate system and
bath at $x=0$.  The cavity thus runs from $x=-l$ to $x=0$, and the
channel region from $x=0$ to $\infty$. The boundary conditions at the
interface are only restricted by the requirement that they lead to an
Hermitian eigenvalue problem.  Below we address two different such
boundary conditions: In the first case we set Neumann boundary
conditions for the cavity; Hermiticity \cite{Savi03,Vivi03} then
imposes Dirichlet conditions for the channel problem. In the second
case we consider the inverse situation with Dirichlet boundary
condition for the cavity and Neumann conditions outside.

\subsection{Cavity with von--Neumann boundary conditions}
\label{sec:CN-CD}

The inside/outside decomposition of the differential operator $L$
reads \cite{Vivi03}
\numparts
\label{eq:Lproj01}
\begin{eqnarray}
\label{eq:Lqq01}
L_{\mathcal{QQ}} \mu(x) &= -\frac{1}{n^2}\frac{\rmd^2}{\rmd x^2}\mu(x) +
\frac{\delta(x-0_{-})}{n^2} \frac{\rmd}{\rmd x^{\prime}}\mu(x^\prime)
\biggr\vert_{x^{\prime}=0_{-}},\\
\label{eq:Lpp01}
L_{\mathcal{PP}} \nu(x) &= -\frac{\rmd^2}{\rmd x^2}\nu(x) -
\delta^{\prime}(x-0_{+}) \nu(0_{+}),\\
\label{eq:Lpq01}
L_{\mathcal{PQ}} \mu(x) &= \frac{\delta^{\prime}(x-0_{+})}{n}
\mu(0_{-}),\\
\label{eq:Lqp01}
L_{\mathcal{QP}} \nu(x) &= -
\frac{\delta(x-0_{-})}{n}\frac{\rmd}{\rmd x^{\prime }}\nu(x^\prime)
\biggr\vert_{x^{\prime}=0_{+}}.
\end{eqnarray}
\endnumparts
The shorthands $0_{\mp}$ indicate the limits where the interface at
$x=0$ is approached from inside respectively outside the
resonator. The singular terms guarantee the matching conditions for
the electromagnetic field at the interface. In addition, these terms
ensure the Hermiticity of $L_{\mathcal{QQ}}$ and
$L_{\mathcal{PP}}$. The range of the operators in
equations (\ref{eq:Lqq01}) and (\ref{eq:Lpp01}) within Hilbert space is
given by the functions for which the singular term vanishes.

The eigenmodes $\mu_{\lambda}(x)$ of the closed cavity are the
solutions of the eigenvalue problem $L_{\mathcal{QQ}}
\mu_{\lambda}(x) = k^2_{\lambda} \mu_{\lambda}(x)$ with
$k_{\lambda}= \omega_{\lambda} /c$. From equation \eref{eq:Lqq01},
they satisfy the equation
\begin{equation}
\label{eq:helmres01}
\frac{\rmd^2}{\rmd x^2} \mu_{\lambda}(x) + n^2 k^2_{\lambda} \mu_{\lambda}(x)
= 0,
\end{equation}
subject to the boundary conditions
\numparts
\begin{eqnarray}
\mu_{\lambda}(-l) = 0 , \\
\frac{\rmd}{\rmd x}\mu_{\lambda}(x) \biggr\vert_{x=0_{-}} = 0.
\end{eqnarray}
\endnumparts
The first condition is imposed by the perfectly reflecting mirror at
$x= -l$, and the second follows from the requirement that the singular
boundary term applied to $\mu_\lambda$ must vanish. The normalized
solutions of the eigenvalue problem form the discrete set
\begin{equation}
\label{eq:ex01mu}
\mu_{\lambda}(x) = \sqrt{\frac{2}{l}} \sin(n k_{\lambda}(x+l)),
\end{equation}
with wave numbers $k_{\lambda} = (2\lambda +1)\pi / 2 n l  \;
(\lambda = 0,1,2,\dots)$. In the channel region, the eigenvalue problem
reads
\begin{equation}
\label{eq:helmbath01}
\frac{\rmd^2}{\rmd x^2} \nu (k,x) + k^2 \nu (k,x) = 0, \quad (k=\omega/c),
\end{equation}
with Dirichlet conditions at the resonator surface,
\begin{equation}
\nu (k,0_{+}) = 0.
\end{equation}
This determines a continuous set of $\delta$--normalized channel
modes,
\begin{equation}
\label{eq:ex01nu}
\nu (k,x) = \sqrt{\frac{2}{\pi}} \sin(k x).
\end{equation}
Since both the cavity and channel modes are real valued functions,
the coupling amplitudes $\mathcal{W}$ and $\mathcal{V}$ become real
and identical, $\mathcal{W}=\mathcal{V}$. Combining equations
(\ref{eq:coupw}), (\ref{eq:Lqp01}) with the mode functions
(\ref{eq:ex01mu}) and (\ref{eq:ex01nu}), we obtain
\begin{equation}
\label{eq:W01}
\mathcal{W}_{\lambda}(k) = \mathcal{V}_{\lambda}(k)
= \frac{(-1)^{\lambda+1}}{n}
\sqrt{\frac{k}{\pi k_{\lambda} l}}.
\end{equation}
The result for the internal frequencies $\omega_{\lambda} = c
k_{\lambda}$, together with the coupling amplitudes (\ref{eq:W01}),
and the mode functions (\ref{eq:ex01mu}) and (\ref{eq:ex01nu}),
completely specify the system--and--bath Hamiltonian and the electric
and magnetic field.

\begin{figure}[t]
  \begin{center} \includegraphics[scale=1]{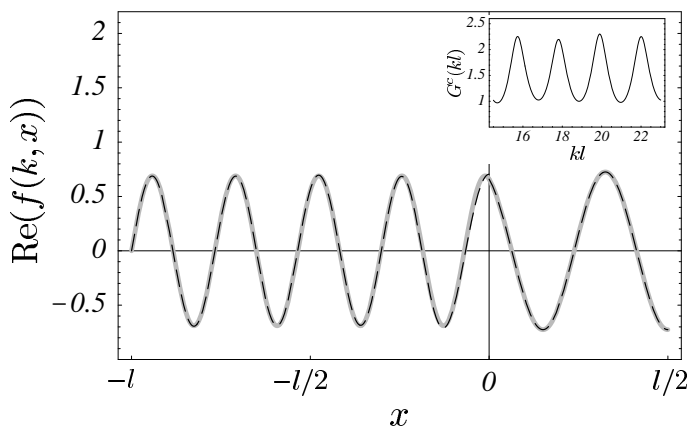}
    \caption{\label{fig:diel1d01} Real part of the scattering wave
    function corresponding to $kl=18$, computed for a one dimensional
    dielectric cavity with refractive index $n=1.5$. The solid line is
    the exact solution, the dashed line the expansion in terms of the
    resonator and channels modes. Only $11$ cavity modes with
    $k_{\lambda}l$ around $kl=18$ were included. The inset shows the
    cavity gain factor as function of $kl$ for a range around
    $kl=18$.}
\end{center}
\end{figure}

We now turn to an illustration of our results. To compare the exact
scattering states with their representation in terms of cavity and
channel modes, we combine equations \eref{eq:expcoefa},
\eref{eq:expcoefb} with the results (\ref{eq:f01}), (\ref{eq:ex01mu}),
and (\ref{eq:ex01nu}). This yields the expansion coefficients
\begin{eqnarray}
\label{eq:a01}
\alpha_{\lambda}(k) = \frac{(-1)^{\lambda + 1} I_k k \cos(n k
  l)}{n\sqrt{\pi l} (k^2-k_{\lambda}^2)} , \\
\label{eq:b01}
\beta(k,k') =
  \frac{1}{2\pi} \left[ \mathcal{P} \left( \frac{2k' (1+S_k)}{k'^2 -
  k^2} \right) - \rmi \pi (1-S_k) \delta(k' - k) \right] .
\end{eqnarray}
The symbol $\mathcal{P}$ denotes the principal value. The S-matrix
$S_k$ and the amplitude $I_k$ are given, respectively, by equations
(\ref{eq:s01}) and (\ref{eq:I01}). \Fref{fig:diel1d01} shows the real
part of the scattering wave function with wavenumber $kl=18$. We
compare the exact result (solid gray line) with the system--and--bath
expansion (dashed line). In the resonator region we only included 11
cavity modes with wavenumber centered around $kl=18$.  The agreement
is very good; deviations are only visible close to $x=0$, i.e.\ near
the boundary separating system and bath.  It has been argued before
\cite{Barn88} that cavity or channel expansions must fail close to the
boundary; so a remark concerning the status of such expansions is in
order here: The inclusion of all cavity and all channel modes yields
an exact point--to--point representation of the scattering function
and its derivative, everywhere {\em except} for the point $x=0$.  This
representation does not converge uniformly but it is exact in the
$L^2$ sense. Therefore the system--and--bath expansion is an exact
representation of the scattering state in the underlying Hilbert
space.

To determine the cavity resonances we solve the eigenvalue problem for
the non-Hermitian operator $L_{\rm eff}(k)$. Explicit calculation
(\ref{app:appB}) shows that $L_{\rm eff}(k)$ acts on an
arbitrary resonator state $\mu(x)$ according to
\begin{equation}
\label{eq:Leff01b}
L_{\rm eff}(k) \mu(x) = -\frac{1}{n^2}\frac{\rmd^2}{\rmd x^2}\mu(x)
  + \frac{\delta(x-0_{-})}{n^2} \left[ \frac{\rmd}{\rmd
  x^{\prime}}\mu(x^\prime) \biggr\vert_{x^{\prime}=0_{-}} - \rmi k
  \mu(0_{-}) \right].
\end{equation}
Due to the singular term, the action of $L_{\rm eff}(k)$ generally
goes beyond Hilbert space. The range of $L_{\rm eff}(k)$ within
Hilbert space is defined by the wave functions for which the singular
term vanishes. It follows that the right eigenstates $\xi_{j}(k,x)$
are solutions of the Helmholtz equation
\begin{equation}
\label{eq:leff01evq}
\frac{\rmd^2}{\rmd x^2} \xi_{j}(k,x) + n^2 \sigma_{j}^2(k) \xi_{j}(k,x) = 0,
\end{equation}
that obey the boundary conditions
\numparts
\begin{eqnarray}
\xi_{j}(k,-l) = 0, \\
\label{eq:bcres01}
\frac{\rmd}{\rmd x^{\prime}}\xi_{j}(k,x^\prime)
\biggr\vert_{x^{\prime}=0_{-}} = \rmi k \xi_{j}(k,0_{-}).
\end{eqnarray}
\endnumparts
The first condition results from the perfect mirror at $x=-l$ while
the second defines the so--called Siegert boundary condition. It
accounts for the leakage out of the cavity: In the channel region that
boundary condition implies a purely outgoing wave. For a fixed value
of $k$, one finds the discrete set of solutions
\begin{equation}
\label{eq:rres01}
\xi_{j}(k,x) = A_{j}(k) \sin(n \sigma_{j}(k)(x+l)) ,
\end{equation}
with some normalization factors $A_{j}(k)$. Substituting the solutions
into equation \eref{eq:bcres01} we obtain the secular equation for the
eigenvalues,
\begin{equation}
\label{eq:res01}
\sigma_{j}(k) = \frac{\rmi}{n} k \tan(n \sigma_{j}(k) l).
\end{equation}
The fixed point equation $k = \sigma_{j}(k)$ determines the cavity
resonances. Analytical continuation of equation \eref{eq:res01} into
the complex plane then yields the resonance condition
\begin{equation}
\label{eq:rescond01}
\tan(n k l) + \rmi n = 0,
\end{equation}
which has solutions only for complex $k$. The resonances can be found
analytically and are given by
\begin{equation}
\label{eq:resex01}
k_j = \frac{1}{n l}\cases
         {\frac{(2j+1)\pi}{2} + \frac{\rmi}{2}
\ln\left(|r|\right);\quad j=0,1,\dots  & $(n > 1)$, \\
         j\pi + \frac{\rmi}{2}
\ln\left(|r|\right);\quad j=1,2,\dots &  $(n < 1)$,}
\end{equation}
where $r = (n-1)/(n+1)$ is the reflection amplitude at the dielectric
surface. Comparison with the direct calculation (cf.\ equation
\eref{eq:s01}) shows that the resonances coincide with the poles of
the scattering--matrix. All resonances have the same width and are
located along a straight line in the lower half of the complex
plane. The resonance spacing, i.e.\ the difference in real parts of
two successive resonances, is constant, $\Delta =\pi/nl$. The
resonances start to overlap when the modulus of the reflection
amplitude becomes smaller than $|r| = \exp (-\pi)$.

We finally evaluate the cavity gain factor. The free--space local
density of states is $\rho_{0}(k,x) = \sqrt{\frac{2}{\pi}} \sin^2
(k(x+l))$. Integration over the cavity volume yields
\begin{equation}
\label{eq:ldos1d}
\int\limits_{-l}^{0} \! \rmd x \, \rho_{0}(k,x) = \frac{l}{\pi}
\left[ 1 - \frac{\sin(2kl)}{2kl} \right].
\end{equation}
The integrated cavity density of states follows from equations
(\ref{eq:aldos}) and (\ref{eq:a01}) by means of the Poisson sum
rule,
\begin{eqnarray}
\label{eq:ldoscav01}
\int\limits_{-l}^{0} \! \rmd x \, \rho(k,x) &= \frac{|I_{k}|^2
  k^2 \cos^2(nkl)}{\pi l n^2} \sum\limits_{\lambda=0}^{+\infty}
\frac{1}{(k^2 - k_{\lambda}^2)^2} \nonumber\\
& = \frac{l|I_{k}|^2}{4\pi} \left[1 - \frac{\sin(2nkl)}{2nkl}\right].
\end{eqnarray}
Combining these results with equation \eref{eq:I01} for $I_k$, we
obtain the cavity gain factor
\begin{equation}
\label{eq:cgf01}
G^{c}(k) = \frac{n^2
\left[1-\frac{\sin(2nkl)}{2nkl}\right]}{\left[n^2 \cos^2(nkl) +
\sin^2(nkl)\right]\left[1-\frac{\sin(2kl)}{2kl} \right]}.
\end{equation}
The inset in \fref{fig:diel1d01} shows the cavity gain factor over a
range of $kl$. The peaks are equally spaced and have approximately the
same high and width as expected from equation \eref{eq:resex01}.

\subsection{Cavity with Dirichlet boundary conditions}
\label{sec:CD-CN}

It is interesting to carry out the system--and--bath quantization in a
basis other than that considered in the previous section.  To that end
we reconsider the dielectric resonator of \fref{fig:cavities}(a)
but perform the system--and--bath quantization with
\textit{interchanged} boundary conditions at the resonator/channel
interface: The resonator modes are now required to satisfy Dirichlet
boundary conditions at $x=0$ while Neumann conditions hold for the
channel modes. The differential operators corresponding to this choice
have the form
\numparts
\label{eq:Lproj02}
\begin{eqnarray}
\label{eq:Lqq02}
L_{\mathcal{QQ}} \mu(x) &= -\frac{1}{n^2}\frac{\rm\rmd^2}{\rmd x^2}\mu(x) +
\frac{\delta^{\prime}(x-0_{-})}{n^2} \mu(0_{-}) \, ,\\
\label{eq:Lpp02}
L_{\mathcal{PP}} \nu(x) &= -\frac{\rm\rmd^2}{\rmd x^2}\nu(x) - \delta(x-0_{+})
\frac{\rmd}{\rmd x^{\prime}}\nu(x^\prime) \biggr\vert_{x^{\prime}=0_{+}} \, ,\\
\label{eq:Lpq02}
L_{\mathcal{PQ}} \mu(x) &= \frac{\delta(x-0_{+})}{n}
\frac{\rmd}{\rmd x^{\prime }}\mu(x^\prime) \biggr\vert_{x^{\prime}=0_{-}}\, ,\\
\label{eq:Lqp02}
L_{\mathcal{QP}} \nu(x) &= -\frac{\delta^{\prime}(x-0_{-})}{n}
\nu(0_{+}).
\end{eqnarray}
\endnumparts The closed cavity eigenmodes of $L_{\mathcal{QQ}}$ solve
the Helmholtz equation and satisfy Dirichlet boundary conditions both
at $x=l$ and $x=0_-$. The second of these conditions follows from the
requirement that the application of $L_{\mathcal{QQ}}$ on any
eigenmode must yield a vanishing singular contribution. The eigenmodes
form the discrete set of functions
\begin{equation}
\label{eq:ex02mu}
\mu_{\lambda}(x) = \sqrt{\frac{2}{l}} \sin(n k_{\lambda}(x+l)),
\end{equation}
with eigenvalues $k_{\lambda} = \pi\lambda/n l \; (\lambda =
1,2,\dots)$. In a similar fashion one finds the continuous set of
channel modes
\begin{equation}
\label{eq:ex02nu}
\nu(k,x) = \sqrt{\frac{2}{\pi}} \cos(k x),
\end{equation}
that satisfy Neumann boundary conditions at $x=0_+$.  Substituting the
mode functions into the definitions (\ref{eq:coupw}) and
(\ref{eq:coupv}) we obtain the coupling amplitudes
\begin{equation}
\label{eq:W02}
\mathcal{W}_{\lambda k} = \mathcal{V}_{\lambda k} =
 (-1)^{\lambda} \sqrt{\frac{k_{\lambda}}{\pi k l}}.
\end{equation}
We note that the cavity eigenfrequencies and the coupling amplitudes
obtained with the present set of boundary conditions differ from the
results obtained in the previous section. Consequently, two different
system--and--bath Hamiltonians are obtained in the two cases. However,
as we show below both Hamiltonians provide an equivalent, and exact,
description of the field dynamics.

\begin{figure}[t]
  \begin{center} \includegraphics[scale=1]{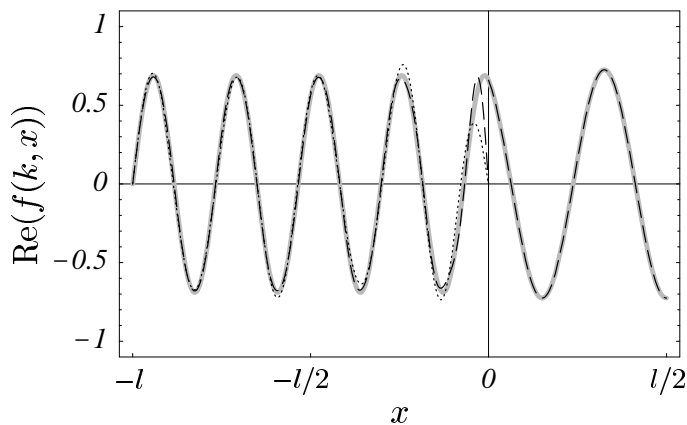}
    \caption{\label{fig:diel1d02} Real part of the scattering wave
      function for a one dimensional dielectric cavity with the
      same parameters as in \fref{fig:diel1d01}. The solid gray
      curve is the exact solution. The system--and--bath expansion
      (dotted, dashed line) is based on $11$, respectively, $25$
      cavity modes satisfying Dirichlet boundary conditions at
      $x=0$.  The dashed line for $x>0$ is the representation in
      terms of channel modes.}
\end{center}
\end{figure}

Expanding the modes--of--the--universe $f(k,x)$ in terms of resonator
and channel modes, we find the expansions coefficients
\begin{eqnarray}
\label{eq:a02}
\alpha_{\lambda}(k) = \frac{(-1)^{\lambda} I_k k_{\lambda}\sin(n k
l)}{n \sqrt{\pi l} (k^2-k_{\lambda}^2) }, \\
\label{eq:b02}
\beta(k,k') = \frac{\rmi}{2\pi} \left[ \mathcal{P} \left( \frac{2k
(1-S_k)}{k'^2 - k^2} \right) - \rmi \pi (1+S_k) \delta(k' - k) \right],
\end{eqnarray}
with $S_k$ and $I_k$ defined, respectively, by equations \eref{eq:s01}
and \eref{eq:I01}. In \fref{fig:diel1d02} we compare the exact
scattering wave function (\ref{eq:f01}) (solid gray line) with the
mode expansion in terms of the expansion coefficients (\ref{eq:a02})
and (\ref{eq:b02}). Perfect agreement is found in the channel region
$x > 0$. In the cavity region there is slow convergence close to
$x=0$, due to the Dirichlet boundary conditions at $x=0_-$. The slower
convergence visible in \fref{fig:diel1d02} must be compared with the
faster convergence found for the other set of boundary conditions
(\Fref{fig:diel1d01}). It indicates that, in spite of the freedom
inherent in the projection formalism for the choice of boundary
conditions, certain boundary conditions are better suited for the
problem yielding good approximations with less terms in the mode
expansions.

As in \sref{sec:CN-CD} we can now evaluate the system resonances and
the cavity gain factor. With the present boundary conditions the
operator $L_{\rm eff}(k)$ reduces to (see \ref{app:appB})
\begin{equation}
\label{eq:Leff02b}
L_{\rm eff}(k) \mu(x) = -\frac{1}{n^2}\frac{\rm\rmd^2}{\rmd x^2}\mu(x) +
  \frac{\delta'(x-0_{-})}{n^2} \left[ \mu(0_{-}) + \frac{\rmi}{k}
\frac{d}{\rmd x'} \mu(x') \biggr\vert_{x^{\prime}=0_{-}} \right].
\end{equation}
It is illustrative to compare this with the result (\ref{eq:Leff01b})
that holds for interchanged resonator/channel boundary
conditions. Both results differ in their singular terms.  However,
upon projection onto the Hilbert space the same operator is recovered
as the singular contributions vanish. In both cases the resulting
boundary condition at the resonator/channel--interface is the Siegert
condition (\ref{eq:bcres01}).

The integrated cavity density of states follows upon combination of
equations (\ref{eq:aldos}) and (\ref{eq:a02}), with the result
\begin{equation}
\label{eq:ldoscav02}
\int\limits_{-l}^{0} \! \rmd x \, \rho(k,x) = 
\frac{l|I_{k}|^2}{4\pi} \left[1 - \frac{\sin(2nkl)}{2nkl}\right].
\end{equation}
It agrees with the result (\ref{eq:ldoscav01}) obtained for the
other set of boundary conditions. This demonstrates that the
physical observables are indeed independent of the choice of
boundary conditions.

\section{One dimensional cavity with a semitransparent mirror}
\label{sec:ex02}

The model of Ley and Loudon \cite{Ley87} is a one dimensional cavity
defined by a totally reflecting mirror at one end and a
semitransparent mirror at the other end (\Fref{fig:cavities}(b)). The
electric field is linearly polarized in the $z$--direction. Radiation
can leak out through the semitransparent mirror modeled by a
dielectric slab of width $d$ and refractive index $n$. The limit $d
\to 0$ and $n \to \infty$ with $n^2 d=\eta$ fixed is taken at the end
of the calculation, here $\eta$ is a factor characterizing the mirror
transparency. In this limit, the frequency dependent mirror reflection
and transmission amplitudes are given by
\begin{equation}
\label{eq:rt}
r(k) = \frac{\rmi k\eta}{2 - \rmi k\eta}, \qquad t(k) =
\frac{2}{2 - \rmi k\eta}.
\end{equation}
They obey the common relations for symmetric mirrors, $|r|^2 + |t|^2
=1$ and $rt^{*} + r^{*}t = 0$.

\begin{figure}[t]
  \begin{center} \includegraphics[scale=1]{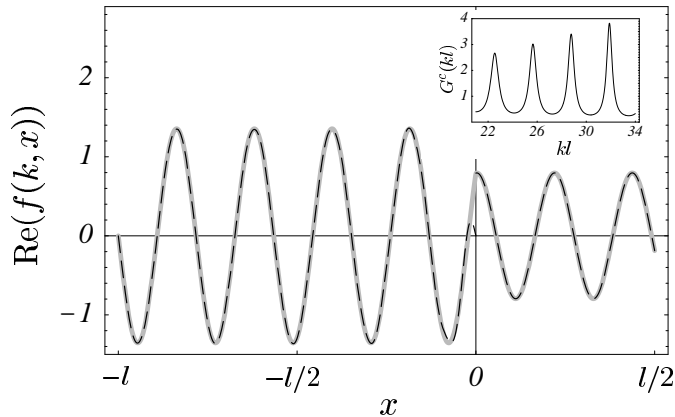}
    \caption{\label{fig:smt1d} Real part of the scattering wave
      function with $kl=28.9$ for a one dimensional optical cavity
      with a perfectly reflecting mirror at $x=-l$ and a
      semitransparent mirror at $x=0$. The mirror transparency is
      characterized by $\eta=0.0453$ corresponding to the reflection
      coefficient $|r(kl=28.9)|^2=0.3$. The solid line is the exact
      solution, the dashed line represents the mode expansion
      truncated to the first 35 modes in the cavity region. Inset:
      Cavity gain factor as function of $kl$ for a range around
      $kl=28$.} \end{center}
\end{figure}

The exact eigenmodes of Maxwell's equations for this problem are
given in equation \eref{eq:f03}. Within the system--and--bath approach
there are two natural ways of a resonator/channel separation:
Either one assumes the mirror to be part of the cavity or the
mirror is part of the channel region. Here, we stick to the latter
choice. Accordingly, the cavity runs from $x=-l$ to $x=0$ and the
channel region from $x=0$ to $\infty$. The alternative definition
with the mirror being part of the cavity can easily be shown to
lead to the same physical results. We choose Dirichlet conditions
for the resonator boundary at $x=0_{-}$, which implies von Neumann
conditions for the outside problem. The differential operators
corresponding to these definitions are
\numparts
\label{eq:Lproj03}
\begin{eqnarray}
\label{eq:Lqq03}
L_{\mathcal{QQ}} \mu(x) &= -\frac{\rmd^2}{\rmd x^2}\mu(x) +
\delta'(x-0_{-}) \mu(0_{-}),\\
\label{eq:Lpp03}
L_{\mathcal{PP}} \nu(x) &=
-\frac{1}{n(x)}\frac{\rmd^2}{\rmd x^2}
\left(\frac{\nu(x)}{n(x)}\right) - \frac{\delta(x-0_{+})}{n(x)}
\frac{\rmd}{\rmd x'}\left(\frac{\nu(x')}{n(x')}\right) \biggr\vert_{x'=0_{+}},\\
\label{eq:Lpq03}
L_{\mathcal{PQ}} \mu(x) &= \frac{\delta(x-0_{+})}{n(x)}
\frac{\rmd}{\rmd x'}\mu(x') \biggr\vert_{x'=0_{-}},\\
\label{eq:Lqp03}
L_{\mathcal{QP}} \nu(x) &= -\delta'(x-0_{-})
\frac{\nu(0_{+})}{n(0_{+})},
\end{eqnarray}
\endnumparts
where $n(x)$ is the refractive index in the channel region,
\begin{equation}
n(x) = \cases{ n & $(0_{+} \le x \le d)$, \\
                     1 & $(d < x)$.}
\end{equation}
The eigenvalue problem defined by $L_{\mathcal{QQ}}$ reduces to that
of the dielectric resonator of \sref{sec:CD-CN} in the case when
the dielectric function equals $1$. Adopting our earlier results in
that limiting case we find the closed cavity eigenmodes
\begin{equation}
\label{eq:ex03mu}
\mu_{\lambda}(x) = \sqrt{\frac{2}{l}} \sin(k_{\lambda}(x+L)),
\end{equation}
with the eigenvalues $k_{\lambda}=\pi \lambda /l \; (\lambda
=1,2,\dots)$. The channel modes are the solutions of the Helmholtz
equation
\begin{equation}
\label{eq:helm03}
\frac{\rmd^2}{\rmd x^2} \nu(k,x) + n^2(x) k^2 \nu(k,x) = 0,
\end{equation}
with Neumann boundary conditions at $x=0_+$. In addition, they must
satisfy the two conditions
\begin{eqnarray}
\label{eq:mirrorBC}
\frac{1}{n} \nu(k,d_{-}) = \nu(k,d_{+}), \\
\frac{1}{n}\frac{\rmd}{\rmd x} \nu(k,x) \biggr\vert_{x=d_{-}} = \frac{\rmd}{\rmd x}
\nu(k,x) \biggr\vert_{x=d_{+}},
\end{eqnarray}
imposed by the continuity of the electric and magnetic field at the
right end of the semitransparent mirror. The shorthands $d_{\pm}$
indicate the limit where $d$ is approached from the left ($d_-$) or
from the right ($d_+$). Solving for $\nu(k,x)$ and taking the limit
$d \to 0$, $n\to \infty$ with $n^2d= \eta$, one obtains the
following continuous set of channel modes,
\begin{equation}
\label{eq:ex03nu}
\nu(k,x) = \frac{1}{\sqrt{2\pi}} \bigl( \rme^{-\rmi kx} + S_{c}(k)
\rme^{\rmi kx}
\bigr), \quad S_{c}(k) = \frac{\rmi-\eta k}{\rmi+\eta k}.
\end{equation}
The coupling amplitudes follow upon substituting the wavefunctions
(\ref{eq:ex03mu}), (\ref{eq:ex03nu}) into the definitions
(\ref{eq:coupw}) and (\ref{eq:coupv}). The result is
\begin{equation}
\mathcal{W}_{\lambda}(k) = \mathcal{V}_{\lambda}(k) =
\frac{(-1)^{\lambda}}{1-i\eta k} \sqrt{\frac{k_{\lambda}}{\pi k l}}.
\end{equation}
Finally, the representation of the exact modes $f(k,x)$ in terms of
the system and bath modes yields the expansion coefficients
\begin{eqnarray}
\label{eq:a03}
\fl \alpha_{\lambda}(k) = \frac{(-1)^{\lambda} I_k k_{\lambda}\sin(n k
l)}{\sqrt{\pi l} (k^2-k_{\lambda}^2) }, \\
\label{eq:b03}
\fl \beta(k,k') = \frac{\rmi}{2\pi} \left[ \mathcal{P} \left( \frac{1 -
S_{c}^{*}(k') S_{k}}{k' - k} + \frac{S_{c}^{*}(k') - S_{k}}{k'+k}
\right) - \rmi \pi (1+S_{c}^{*}(k)S_k) \delta(k' - k) \right],
\end{eqnarray}
where $S_{k}$ and $I_{k}$ are given by equation \eref{eq:s03} and
equation \eref{eq:I03}, respectively. \Fref{fig:smt1d} shows the real part of
the scattering wave function with $kl=28.9$. The exact solution
(\ref{eq:f03}) (solid gray line) is compared with the mode expansion
(dashed line) using the expansion coefficients (\ref{eq:a03}) and
(\ref{eq:b03}). The first 35 cavity modes were included. The
deviations from the exact scattering wave function visible near $x=0$
can be made arbitrary small by including more terms in the mode
expansion.

In order to find the resonances one must solve the eigenvalue equation
for $L_{\rm eff}(k)$. The real space representation of $L_{\rm
  eff}(k)$ for our choice of boundary conditions follows upon
combination of equations \eref{eq:Leff} and (\ref{eq:Lqq03}) (see
\ref{app:appB}), with the result
\begin{equation}
\label{eq:Leff03b}
\fl L_{\rm eff}(k) \mu(x) = -\frac{\rmd^2}{\rmd x^2}\mu(x) + \delta'(x-0_{-})
\left[ \mu(0_{-}) - \frac{1}{k (\rmi +\eta k)} \frac{\rmd}{\rmd x'} \mu(x')
  \biggr\vert_{x^{\prime}=0_{-}} \right],
\end{equation}
where $\mu(x)$ is an arbitrary resonator state. The range of $L_{\rm
eff}(k)$ within Hilbert space is defined by the resonator functions
for which the singular term on the right hand side vanishes. In
particular, this holds for the right eigenstates $\xi_{j} (k,x)$. It
follows that these states satisfy the boundary condition
\begin{equation}
\label{eq:bcres03}
\xi_{j}(k,x) = \frac{1}{k(\rmi +\eta k)}
\frac{\rmd}{\rmd x}\xi_{j}(k,0_{-}) \biggr\vert_{x=0_{-}} \, ,
\end{equation}
at the semitransparent mirror. There is a discrete set of solutions,
\begin{equation}
\label{eq:rres03}
\xi_{j}(k,x) = A_{j}(k) \sin(\sigma_{j}(k)(x+L)),
\end{equation}
with some normalization constant $A_{j}(k)$. Substitution into
equation \eref{eq:bcres03} yields the equation for the eigenvalues
$\sigma_{j}(k)$,
\begin{equation}
\label{eq:res03}
\sigma_{j}(k) \cot(\sigma_{j}(k) l) = k(\rmi + k\eta).
\end{equation}
After analytical continuation of the fixed point equation $k =
\sigma_{j} (k)$ into the complex plane we obtain the resonance
condition
\begin{equation}
\label{eq:rescond03}
\rmi + \eta k - \cot(k l) = 0 \, , 
\end{equation} 
that coincides with the equation for the poles of the $S$-matrix (cf.\
equation \eref{eq:s03}).

To quantify the resonant response of the cavity to external
excitations we compute the integrated local density of states, again
using the Poisson sum rule,
\begin{eqnarray}
\label{eq:ldoscav03}
\int\limits_{-l}^{0} \! \rmd x \, \rho(k,x) &= \frac{l|I_{k}|^2
 \sin^2(kl)}{l^2\pi} \sum\limits_{\lambda=1}^{+\infty}
\frac{k_{\lambda}^2}{(k^2 - k_{\lambda}^2)^2}, \nonumber\\
& = \frac{l|I_{k}|^2}{4\pi} \left[1 - \frac{\sin(2kl)}{2kl}\right].
\end{eqnarray}
Combination with the free--space density of states (\ref{eq:ldos1d})
gives the cavity gain factor
\begin{equation}
\label{eq:cgf03}
G^{c}(k) =
\frac{1}{\left[1- \eta k \sin(2kl) + (\eta k)^2 \sin^2(nkl)\right]}.
\end{equation}
With increasing $kl$ sharper resonances are found in the cavity
gain factor (see \fref{fig:smt1d}(inset)). The reason is the
reduction of the mirror transmission for large $kl$ that, in turn,
enhances the lifetime of the cavity resonances.

\section{Dielectric disk}
\label{sec:ex04}

In this section we demonstrate our quantization technique for
resonators of spatial dimension larger than one. Specifically, we
consider a two--dimensional circular dielectric of radius $R$ and
refractive index $n$ (see \fref{fig:cavities}(c)). The resonator
is embedded in free space.  We restrict ourselves to TM modes with the
electric field polarized in the $z$--direction. It is convenient to
use polar coordinates ${\bm r} = (r,\phi)$ below. The dielectric
function then reads
\begin{equation}
\epsilon (r) = n^2 \Theta (R-r) + \Theta (r - R).
\end{equation}
The scattering problem at the resonator can be solved exactly. The
exact eigenmodes of Maxwell's equations are summarized in equation
\eref{eq:f04}.

To apply our quantization technique we separate system and bath along
the boundary of the dielectric: The dielectric disk ($r \le R_{-}$) is
taken as the cavity, while the free space ($r \ge R_{+}$) becomes the
channel region.  For the cavity we assume Dirichlet boundary
conditions at $r=R_{-}$, which implies Neumann conditions at $r=R_{+}$
for the channel problem. The differential operator resulting from this
choice reads
\numparts
\label{eq:Lproj04}
\begin{eqnarray}
\label{eq:Lqq04}
L_{\mathcal{QQ}} \mu(r,\phi) &= -\frac{1}{n^2} \nabla^2 \mu(r,\phi) +
\frac{\partial}{\partial r} \left( \frac{1}{r} \delta (r-R_{-})
\right) \frac{R_{-}}{n^2} \mu(R_{-},\phi),\\
\label{eq:Lpp04}
L_{\mathcal{PP}} \nu(r,\phi) &= -\nabla^2 \nu(r,\phi) -
\delta(r-R_{+}) \frac{\partial}{\partial r'} \nu(r',\phi) \biggr\vert_{r'=R_{+}},\\
\label{eq:Lpq04}
L_{\mathcal{PQ}} \mu(r,\phi) &= \frac{\delta(r-R_{+})}{n}
\frac{\partial}{\partial r'} \mu(r',\phi) \biggr\vert_{r'=R_{-}},\\
\label{eq:Lqp04}
L_{\mathcal{QP}} \nu(r,\phi) &= - \frac{\partial}{\partial r} \left(
\frac{1}{r} \delta (r-R_{-}) \right) \frac{R_{-}}{n} \nu(R_{+},\phi).
\end{eqnarray}
\endnumparts Due to the rotational symmetry we can choose the
eigenstates to be angular momentum eigenstates.

The eigenmodes $\mu_{m\lambda}$ of the closed cavity are labeled by
the angular momentum number $m$ and the radial quantum number
$\lambda$. They solve the Helmholtz equation
\begin{equation}
\label{eq:cav04}
\nabla^2 \mu_{m\lambda}(r,\phi) = - n^2k_{m\lambda}^2
\mu_{m\lambda}(r,\phi),
\end{equation}
and satisfy the Dirichlet condition $\mu_{m\lambda}(R_{-},\phi) =
0$. The normalized eigenstates are given in terms of Bessel functions
of the first kind,
\begin{equation}
\label{eq:mu04}
\mu_{m\lambda}(r,\phi) = \frac{\rme^{\rmi m\phi}
J_{m}(nk_{m\lambda}r)}{\sqrt{\pi}R J_{m+1}(x_{m\lambda})} ,
\end{equation}
with $m = 0,\pm 1, \pm2, \dots$ and $\lambda = 0,1,2,\dots$. The
eigenvalues are $k_{m\lambda} = x_{m\lambda}/ nR$ where
$x_{m\lambda}$ denotes the $\lambda$-th zero of $J_{m}(r)$. In a
similar fashion, one determines the eigenstates in the channel
region. They can be written in terms of Hankel functions,
\begin{equation}
\label{eq:nu04}
\nu_{m}(k,r,\phi) = \sqrt{\frac{k}{8\pi}} \rme^{\rmi m\phi} (H_{m}^{(2)}(kr) +
S_{m}(k) H_{m}^{(1)}(kr)),
\end{equation}
with the diagonal element of the scattering matrix
\begin{equation}
S_{m}(k) = - \frac{H_{m}^{\prime (2)}(kR)}{H_{m}^{\prime (1)}(kR)}.
\end{equation}
The channel states obey the Neumann condition
$\frac{\partial}{\partial r}\nu_{m} (k,r,\phi) \Bigr \vert_{r=R_{+}} =
0$.

In addition to the internal frequencies $\omega_{m\lambda} =
ck_{m\lambda}$ we need the coupling amplitudes $\mathcal{W}$ and
$\mathcal{V}$ to fully determine the system--and--bath Hamiltonian.
Combination of equation \eref{eq:Lqp04} with the mode functions
(\ref{eq:mu04}), (\ref{eq:nu04}) and the definitions
(\ref{eq:coupw}) and (\ref{eq:coupv}) yields after a short
calculation
\begin{eqnarray}
\mathcal{W}_{m\lambda,n}(k) &= \frac{-\rmi \sqrt{2 k_{m\lambda}}}{\pi kR 
H_{m}^{\prime (1)}(kR)}\, \delta_{m,n}, \\
\mathcal{V}_{m\lambda,n}(k) &= \frac{-\rmi \sqrt{2 k_{m\lambda}}}{\pi kR
H_{m}^{\prime (1)}(kR)}\, \delta_{m,-n}.
\end{eqnarray}
We note that the resonant amplitude $\mathcal{W}$ couples only
cavity and channel modes with the same angular momentum, while the
antiresonant amplitude $\mathcal{V}$ couples modes with opposite
angular momentum. This feature guarantees angular momentum
conservation: The resonant terms $\mathcal{W}_{\lambda m}
a_{m}^\dagger b_{m}$ and $\mathcal{W}_{\lambda m}^* a_{m}
b_{m}^\dagger $ account for the creation of a photon with angular
momentum $m$ and the simultaneous annihilation of a second photon
with the same angular momentum. By contrast, the antiresonant
terms, $\mathcal{V}_{\lambda m} a_{m\lambda}b_{-m}$ and
$\mathcal{V}_{\lambda m}^* a_{m\lambda} b_{-m}$ describe the
simultaneous annihilation or creation of two photons with opposite
value of angular momentum. In both cases, the total angular
momentum is conserved.

\begin{figure}[t]
  \begin{center} \includegraphics[scale=1]{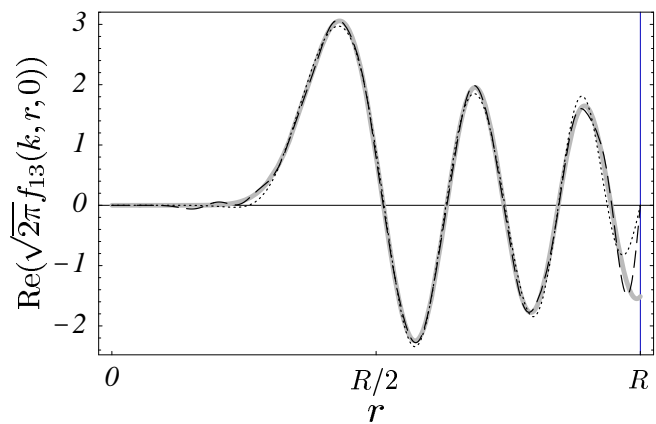}
    \caption{\label{fig:ddk2d} Real part of the radial component of
      the scattering wave function with angular momentum $m=13$ and
      $kR=10.5$ inside a dielectric disk with radius $R$ and index
      of refraction $n=3.3$. The solid gray curve is the exact
      result. The dotted (dashed) line follows from the mode
      expansion taking into account the first 11 (25) cavity
      modes.}
\end{center}
\end{figure}

We now turn to the electromagnetic field and the cavity resonances.
In the cavity region the exact scattering states $f_{m}(k)$ can be
represented in terms of the cavity modes $u_{m\lambda}$, with the
expansion coefficients
\begin{equation}
\label{eq:a04}
\alpha_{m\lambda,n}(k) = \sqrt{\frac{k}{2}} \frac{ k_{m\lambda} I_{mk}
J_{m}(nkR)}{ n (k^2 - k_{m\lambda}^2)} \delta_{m,n},
\end{equation}
where $I_{mk}$ is given by equation \eref{eq:I04}. It suffices to
compare the radial component of the scattering wave functions. In
\fref{fig:ddk2d} we show the real part of that component for
angular momentum $m=13$ and $kR=10.5$. The solid gray line is the
exact result (\ref{eq:f04}) while the dotted and dashed lines
represent the system--and--bath expansion using the first 11 and 25
cavity modes, respectively.

The cavity resonances are obtained by solving the eigenvalue
problem for $L_{\rm eff}(k)$. The calculation is presented in
\ref{app:appB} and yields the resonance condition
\begin{equation}
\label{eq:rescircle}
J_{m}(n k R) H_{m}^{\prime (1)}(kR) - n J_{m}^{\prime}(n k R)
H_{m}^{(1)}(kR) = 0,
\end{equation}
which is equivalent to the equation that determines the poles of
the $S$-matrix (cf.\ equation \eref{eq:s04}).

\begin{figure}[t]
  \begin{center} \includegraphics[scale=1]{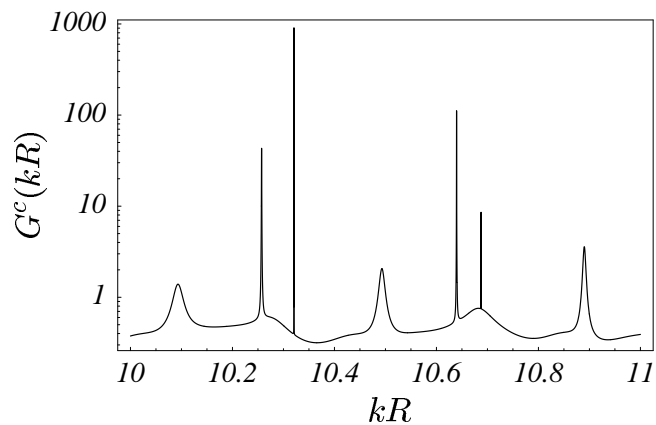}
    \caption{\label{fig:ddcgf} Cavity gain factor as function of $kR$
      for a dielectric disk with radius $R$ and index of refraction
      $n=3.3$. The sharp resonances correspond to states with
      angular momentum $10< m <18$, very sharp resonances with $m
      \gg 18$ are not resolved.}
  \end{center}
\end{figure}

Substitution of equation \eref{eq:a04} into equation \eref{eq:aldos} yields
the integrated local density of states inside the dielectric disk
(see \ref{app:appC}),
\begin{equation}
\label{eq:ldos04}
\fl \int\limits_{\rm disk} d\bm{r} \rho(k,\bm{r}) =
\sum_{m=-\infty}^{+\infty} \frac{k R^2 |I_{mk}|^2}{8 n^2} \left(
J_{m}^{2}(nkR) - J_{m+1}(nkR)J_{m-1}(nkR) \right).
\end{equation}
Together with the free--space local density of states $\rho_{0}(k)
= k/2\pi$, we obtain the cavity gain factor
\begin{equation}
G^{c}(k) = \frac{4 \left(J_{m}^{2}(nkR) - J_{m+1}(nkR)J_{m-1}(nkR)
\right)}{(\pi n k R)^2 | J_{m}(nkR) H_{m}^{\prime(1)} (kR) - n
  J_{m}^{\prime}(nkR) H_{m}^{(1)} (kR) |^2},
\end{equation}
where we used the explicit expression (\ref{eq:I04}) for the mode
amplitude $I_{mk}$. The cavity gain factor displays a set of very
sharp resonances (see \fref{fig:ddcgf}), corresponding to
states with angular momentum $kR < m <nkR$, superimposed over a
smooth background due to broad resonances with $m < kR$
\cite{Hent02}.

\section{Conclusion}
\label{sec:conclusions}

In this paper we have derived exact system--and--bath Hamiltonians
for a number of optical resonators. Physical observables computed
with these Hamiltonian agree with the results obtained by direct
scattering methods. It follows that the system--and--bath
approach, originally developed as a phenomenological model, can
provide an exact quantitative description of optical systems even
in the regime of overlapping modes.

On a formal level our quantization method is based on the Feshbach
projector technique \cite{Fesh62} that has extensively been used in
nuclear and condensed matter physics \cite{Maha69,Ditt00}.  In that
areas the method has become a powerful tool for the description of
disordered and chaotic media.  Statistical theories for such media
have been obtained employing a random--matrix assumption for the
underlying system Hamiltonian \cite{Guhr98}.  Our application of the
projector technique to the field of quantum optics may lay the ground
for a statistical treatment of disordered and wave--chaotic optical
media.

\ack

We thank P.\ Braun, F.\ Haake, D.\ V.\ Savin, H.-J.\ Sommers and J.\
D.\ Urbina for helpful discussions. This work has been supported in
part by an Heisenberg fellowship and by the SFB/TR 12 der Deutschen
Forschungsgemeinschaft.

\appendix

\section{Exact modes of Maxwell's equations}\
\label{app:appA}

In this appendix we summarize the exact solutions of Maxwell's
equations for the three systems treated in the paper; for a more
detailed derivation we refer to references \cite{Ujih75,Ley87,Hent02}.
The modes--of--the--universe are taken to be scattering states with an
incoming wave in only one scattering channel. In the 2d example, this
channel is labeled by the angular momentum number $m$. The scattering
states are normalized according to equation \eref{eq:normuniv}.

\subsection{One dimensional dielectric cavity}
For the dielectric cavity of \fref{fig:cavities}(a) the scattering
states are given by
\begin{equation}
\label{eq:f01}
f(k,x) = \frac{1}{\sqrt{2\pi}} \cases{ 
                \frac{I_{k}}{n} \sin(n k(x+l)) & $(-l < x < 0)$, \\
                \exp(- \rmi kx) + S_k \exp(\rmi kx) & $(0 < x)$.}
\end{equation}
They satisfy the boundary condition $f(k,-l)=0$, imposed by the
completely reflecting mirror at $x=-l$, and are continuous with
continuous derivative for any value of $x>-l$. The
single--channel S--matrix and the mode strength amplitude are
given, respectively, by
\begin{eqnarray}
\label{eq:s01}
S_k = - \frac{n+ \rmi \tan(n k l)}{n- \rmi \tan(n k l)}, \\
\label{eq:I01}
I_k =  \frac{-2\rmi n}{n\cos(nkl) - \rmi \sin(nkl)}\, .
\end{eqnarray}

\subsection{Cavity with semitransparent mirror}
The scattering states for the one dimensional cavity with the
semitransparent mirror (Fig.~\ref{fig:cavities}(b)) have the form
\begin{equation}
\label{eq:f03}
f(k,x) = \frac{1}{\sqrt{2\pi}} 
           \cases{ I_{k}
  \sin(k(x+l)) & $(-l < x < 0)$, \\ 
\exp(-\rmi kx) + S_k \exp(\rmi kx) & $(0 < x)$,} 
\end{equation}
where the scattering matrix $S_k$ and the mode strength amplitude
$I_k$ are given by
\begin{eqnarray}
\label{eq:s03}
S_k =\frac{\rmi -\eta k + \cot(k l)}{\rmi+\eta k - \cot(k l)}, \\
\label{eq:I03}
I_k =\frac{2\rmi}{(\rmi +k\eta)\sin(kl)-\cos(kl)}.
\end{eqnarray}
Here $\eta$ specifies the mirror transparency. The modes satisfy
$f(k,-l)=0$ at the perfectly reflecting mirror and are continuous
everywhere else. At the semitransparent mirror their derivative has
a discontinuity proportional to the mode amplitude, $f'(k,0_{+}) -
f'(k,0_{-}) = -\eta k^2 f(k,0)$, where the prime denotes
differentiation with respect to the position.

\subsection{Dielectric disk}
The exact eigenstates for a two dimensional dielectric disk of
radius $R$ and refractive index $n$ embedded in empty space read
\begin{equation}
\label{eq:f04}
f_m(k,r,\phi) = \sqrt{\frac{k}{8 \pi}} \rme^{- \rmi m\phi} 
           \cases{ I_{mk} J_{m}(nkr) & $(0 < r < R)$, \\ 
H_{m}^{(2)}(kr) + S_{mk} H_{m}^{(1)}(kr) & $(R < r)$,}
\end{equation}
where $m$ labels angular momentum. The channel with index $m$ is
open when $k$ exceeds the channel threshold $k_{m}=m/nR$. Due to
rotational symmetry angular momentum is conserved, and the
$S$--matrix is diagonal in the angular momentum basis
\begin{equation}
\label{eq:s04}
S_{mk} =-\frac{H_{m}^{\prime(2)} (kR) - n [J_{m}^{\prime}(nkR) /
    J_{m}(nkR)] H_{m}^{(2)} (kR)} {H_{m}^{\prime(1)} (kR) - n
  [J_{m}^{\prime}(nkR) / J_{m}(nkR)] H_{m}^{(1)} (kR)}.
\end{equation}
The mode strength amplitude takes the form
\begin{equation}
\label{eq:I04}
I_{mk} =\frac{ 4 \rmi}{\pi k R \left( J_{m}(nkR) H_{m}^{\prime(1)} (kR) - n
  J_{m}^{\prime}(nkR) H_{m}^{(1)} (kR) \right)}.
\end{equation}

\section{External Green functions}
\label{app:appB}

Here we evaluate the action of the non--Hermitian differential
operator $L_{\rm eff}(k)$ on an arbitrary cavity state. According
to equation \eref{eq:Leff} $L_{\rm eff}(k)$ is the sum of two
operators. The first contribution $L_{\mathcal{QQ}}$ has already
been computed in equations (\ref{eq:Lqq01}), (\ref{eq:Lqq02}),
(\ref{eq:Lqq03}) and (\ref{eq:Lqq04}). The second contribution has
the form
\begin{equation}
\label{eq:Leffb}
L_{\mathcal{QP}} \mathcal{G}_{\rm ch}(k) L_{\mathcal{PQ}} \mu(r,\phi)
= L_{\mathcal{QP}} \sum_{m} \int\limits_{k_{m}}^{\infty} \!  \rmd k' \,
\frac{|{\bm \nu}_{m}(k')\rangle \langle {\bm \nu}_{m}(k')|}{k^2 - k'^2
  + \rmi \epsilon} L_{\mathcal{PQ}} \mu(r,\phi) ,
\end{equation}
where $\mathcal{G}_{\rm ch}$ stands for the retarded Green
function of the isolated channel region. Below we compute this
contribution for the cavities of interest.

\subsection{One dimensional channel}

For a one dimensional semi--infinite channel $x \geq 0$ with
Dirichlet boundary conditions at $x=0_+$ we find the retarded Green
function from the solutions (\ref{eq:ex01nu}) by contour
integration,
\begin{eqnarray}
  \mathcal{G}_{\rm ch}(k,x,x') &= \frac{2}{\pi}
  \int\limits_{0}^{\infty} \!
  \rmd k' \frac{\sin(k'x) \sin(k'x')}{k^2 - k'^2 + \rmi \epsilon} \nonumber\\
  &= \frac{\rmi}{2k} \left( \rme^{\rmi k |x+x'|} - \rme^{-\rmi k |x-x'|} \right).
\end{eqnarray}
Combination with the definitions (\ref{eq:Lpq01}), (\ref{eq:Lqp01})
yields the real space representation of $L_{\rm eff}(k)$. The relevant
derivatives have to be done with care as the limits $x \to 0$ and
$x^\prime \to 0$ must be taken independently. The result takes the
form
\begin{equation}
\label{eq:Gex01}
 L_{\mathcal{QP}}
\mathcal{G}_{\rm ch}(k) L_{\mathcal{PQ}} \mu(x) =
-\frac{\rmi k \delta(x-0_{-})}{n^2} \mu(0_{-}).
\end{equation}
Together with equation (\ref{eq:Lqq01}) we arrive at
the result \eref{eq:Leff01b} for $L_{\rm eff}(k)$.

In a similar fashion we evaluate the retarded Green function for
the isolated channel problem with Neumann conditions at $x=0_{+}$.
Using the solutions (\ref{eq:ex02nu}) we obtain in this case
\begin{eqnarray}
\label{eq:Gex02}
\mathcal{G}_{\rm ch}(k,x,x') &= \frac{2}{\pi} \int\limits_{0}^{\infty} \!
\rmd k' \frac{\cos(k'x) \cos(k'x')}{k^2 - k'^2 + \rmi \epsilon} \nonumber\\
&= -\frac{\rmi}{2k} \left( \rme^{\rmi k |x+x'|} + \rme^{-\rmi k |x-x'|} \right).
\end{eqnarray}
Combination with the real space representation (\ref{eq:Lpq02}) and
(\ref{eq:Lqp02}) of $L_{\mathcal{PQ}}$ and $L_{\mathcal{QP}}$,
respectively, yields 
\begin{equation}
\label{eq:Geff02}
 L_{\mathcal{QP}} \mathcal{G}_{\rm ch}(k) L_{\mathcal{PQ}} \mu(x) =
\frac{\rmi \,\delta^\prime (x-0_{-})}{n^2 k} \frac{\rmd}{\rmd x^\prime} 
\mu(x^\prime) \biggl\vert_{x^\prime=0_{-}}.
\end{equation}
Combinations of equations (\ref{eq:Leff}), (\ref{eq:Lqq02}), and
(\ref{eq:Geff02}) yields $L_{\rm eff}(k)$ given in equation
\eref{eq:Leff02b}.

\subsection{Channel with semitransparent mirror}

The semitransparent mirror has width $d$ and refractive index $n$.
In the limit $d \to 0$, $n \to \infty$ with $n^2 d=\eta$ fixed, the
retarded Green function can be expressed in terms of an integral
over products of the scattering states (\ref{eq:ex03nu}),
\begin{equation}
\label{eq:Gex03}
\mathcal{G}_{\rm ch} (k,x,x^\prime) = \frac{1}{2\pi}
\int\limits_{-\infty}^{\infty} \! \rmd k^\prime \frac{\rme^{\rmi k'(x-x')} 
+ S_{c}(k') \rme^{\rmi k'(x+x')}}
{k^2 - k'^2 + \rmi \epsilon} \, .
\end{equation}
The integral can be done by contour integration. Using the unitarity
of $S_{c}(k)$ and taking into account that $S_{c}(k)$ is analytic in
the upper half of the complex plane, equation \eref{eq:Gex03} reduces to
\begin{equation}
\label{eq:Gex03b}
\mathcal{G}_{\rm ch} (k,x,x^\prime) = 
 -\frac{\rmi}{2k} \left( \rme^{\rmi k |x-x^\prime|} + S_{c}(k)
 \rme^{\rmi k |x+x^\prime|}
\right) \, .
\end{equation}
Combination with the definitions (\ref{eq:Lpq03}) and (\ref{eq:Lqp03})
yields
\begin{equation}
\label{eq:Geff03c}
 L_{\mathcal{QP}} \mathcal{G}_{\rm ch}(k) L_{\mathcal{PQ}} \mu(x) =-
\frac{\delta^\prime(x-0_{-})}{k (\rmi + \eta k)} \frac{\rmd}{\rmd x^\prime}
 \mu(x^\prime) \biggl\vert_{x^\prime=0_{-}}.
\end{equation}
Equations (\ref{eq:Lqq03}), (\ref{eq:Geff03c}) along with
equation \eref{eq:Leff} yield equation \eref{eq:Leff03b}.

\subsection{Angular momentum channels}

The retarded Green function for the two--dimensional Helmholtz
equation with Neumann boundary conditions along a disk of radius
$R$, has the form
\begin{eqnarray}
\fl \mathcal{G}_{\rm ch}(k,r',\phi',r,\phi) = - \frac{\rmi}{2\pi}
\sum\limits_{-\infty}^{+\infty} \rme^{\rmi m(\phi-\phi')} \frac{H_{m}^{(1)}
  (kr_{>})}{H_{m}^{\prime (1)} (kR)} \nonumber\\
\times \left( H_{m}^{(2)}
(kr_{<})H_{m}^{\prime (1)} (kR) - H_{m}^{\prime (2)}
(kR) H_{m}^{(1)} (kr_{<}) \right),
\end{eqnarray}
where $r_{<} \; (r_{>})$ stands for the smaller (larger) of $r$ and
$r'$. Substitution of the definitions (\ref{eq:Lpq04}) and
(\ref{eq:Lqp04}) yields
\begin{eqnarray}
\label{eq:Geff04c}
\fl L_{\mathcal{QP}} \mathcal{G}_{\rm ch}(k) L_{\mathcal{PQ}}
\mu(r,\phi) = - \frac{R}{2n^2\pi} \frac{\partial}{\partial r} \left(
\frac{1}{r} \delta (r-R_{-}) \right) \nonumber\\
\times\sum\limits_{m=-\infty}^{+\infty} \int\limits_{0}^{2\pi} \!
d\phi' \; \rme^{\rmi m(\phi-\phi')}
\frac{H_{m}^{(1)}(kR)}{kH_{m}^{\prime (1)}(kR)}
\frac{\partial}{\partial r''}\mu(r'',\phi') \biggr\vert_{r''=R_{-}} \,
.
\end{eqnarray}
Combination with equations (\ref{eq:Leff}) and (\ref{eq:Lqq04}) shows
that $L_{\rm eff}(k)$ acts on an arbitrary resonator state $\mu
(r,\phi)$ like
\begin{eqnarray}
\label{eq:leff04}
\fl L_{\rm eff}(k) \mu(r,\phi) = -\frac{1}{n^2} \nabla^2 \mu(r,\phi) +
\frac{R}{n^2} \frac{\partial}{\partial r} \left( \frac{1}{r} \delta
(r-R_{-}) \right) \Biggl[ \mu(R_{-},\phi) \nonumber\\ - \frac{1}{2\pi}
  \sum\limits_{m=-\infty}^{+\infty} \int\limits_{0}^{2\pi} \! d\phi'
  \; \rme^{\rmi m(\phi-\phi')} \frac{H_{m}^{(1)}(kR)}{kH_{m}^{\prime
      (1)}(kR)} \frac{\partial}{\partial r''}\mu(r'',\phi')
  \biggr\vert_{r''=R_{-}} \Biggr].
\end{eqnarray}
Conservation of angular momentum and the requirement that the singular
terms must vanishes for the right eigenstates $\xi_{mj}(k,r,\phi)$
yield the boundary condition
\begin{equation}
\label{eq:rbcs04}
\xi_{mj}(k,R_{-},\phi) = 
  \frac{H_{m}^{(1)}(kR)}{kH_{m}^{\prime (1)}(kR)}
  \frac{\partial}{\partial r''}\xi_{mj}(k,r'',\phi) \biggr\vert_{r''=R_{-}}.
\end{equation}
There is a discrete set of solutions
\begin{equation}
\label{eq:rres04}
\xi_{mj}(k,r,\phi) = A_{mj}(k) \rme^{\rmi m\phi} J_m(n \sigma_{mj}(k) r),
\end{equation}
with normalization constants $A_{mj}(k)$. Substitution into
equation \eref{eq:rbcs04} yields the equation for the eigenvalues
\begin{equation}
k J_{m}(n \sigma_{mj}(k) R) H_{m}^{\prime (1)}(kR) - n\sigma_{mj}(k)
J_{m}^{\prime}(n \sigma_{mj}(k) R) H_{m}^{(1)}(kR) = 0.
\end{equation}
Combination with the fixed point equation $k = \sigma_{mj}(k)$
finally gives the resonance condition (\ref{eq:rescircle}).

\section{Local density of states inside the dielectric disk}
\label{app:appC}

Here we determine the density of states inside a dielectric disk with
radius $R$ and refractive index $n$. We start from equation
\eref{eq:aldos}. Using the coefficients (\ref{eq:a04}), we obtain
\begin{equation}
\label{eq:C01}
\int\limits_{\rm disk} \rmd\bm{r} \rho(k,\bm{r}) =
\sum_{m=-\infty}^{+\infty} \frac{k |I_{mk}|^2}{2 n^2} J_{m}^{2}(nkR)
\sum\limits_{\lambda=1}^{\infty}\frac{k_{m\lambda}^2}{(k^2-k_{m\lambda}^2)^2}
\, .
\end{equation}
We concentrate on the evaluation of the last sum on the right hand
side. It can be written as a derivative
\begin{equation}
\label{eq:C02}
\sum_{\lambda=1}^{\infty}
\frac{k_{m\lambda}^2}{(k^2-k_{m\lambda}^2)^2} = -\frac{1}{2}
\frac{\partial}{\partial \alpha} \sum_{\lambda=1}^{\infty}
\frac{1}{(k^2-\alpha^2 k_{m\lambda}^2)}
\biggr\vert_{\alpha=1} \, .
\end{equation}
The right hand side can be simplified by noticing that $n R
k_{m\lambda}= x_{m\lambda}$ is the $\lambda$-th zero of the Bessel
function $J_m(x)$. The sum on the right hand side can then be identify
as the trace of the Green function for the radial part of the
Helmholtz equation, in a dielectric disk with Dirichlet conditions on
the disk perimeter. Equation \eref{eq:C02} becomes
\begin{equation}
\label{eq:C03}
\sum_{\lambda=1}^{\infty}
\frac{k_{m\lambda}^2}{(k^2-k_{m\lambda}^2)^2} = -\frac{1}{2}
\frac{\partial}{\partial \alpha} \int\limits_{0}^{R} \! \rmd r \, r
\mathcal{G}_{\rm disk}(k/\alpha,r,r)
\Bigr\vert_{\alpha=1} \, .
\end{equation}
The disk Green function can be found by standard methods
\cite{Jack75} and is given by
\begin{eqnarray}
\label{eq:C04}
\fl \mathcal{G}_{\rm disk}(k/\alpha,r',r) = \frac{\rmi \pi}{4
J_{m}\left(nkR/\alpha \right)} J_{m}\left(nkr_{<}/\alpha \right)
\nonumber \\
\times \left( J_{m}\left(nkR/\alpha \right) Y_{m}\left(nkr_{>}/\alpha
\right) - Y_{m}\left(nk R/\alpha\right) J_{m}\left(nkr_{>}/\alpha
\right) \right),
\end{eqnarray}
where $Y_{m}$ is a Bessel function of the second kind and $r_{<}$
($r_{>}$) the smaller (larger) of $r$ and $r'$. The problem then
reduces to the evaluation of the integral in equation \eref{eq:C03}.
Using the relations
\begin{eqnarray}
\fl \int\limits_{0}^{R} \! \rmd r \,r J_{m}^{2}(kr) = \frac{R^2}{2}
\left( J_{m}^{2}(kR) - J_{m+1}(kR) J_{m-1}(kR) \right) \, , \\ 
\fl
\int\limits_{0}^{R} \! \rmd r \,r J_{m}(kr)Y_{m}(kr)= \frac{R^2}{2}
\Bigl( J_{m}(kR)Y_{m}(kR) - \frac{1}{2} J_{m+1}(kR) Y_{m-1}(kR)
\nonumber\\ - \frac{1}{2} J_{m-1}(kR) Y_{m+1}(kR) \Bigr) \, ,
\end{eqnarray}
we find after some straightforward manipulations the result
\begin{equation}
\int\limits_{0}^{R} \! \rmd r \, r \mathcal{G}_{\rm disk}(k/\alpha,r,r)
= \frac{\alpha R J_{m}^{\prime}(kR/\alpha)}{2kJ_{m}(kR/\alpha)}.
\end{equation}
Substitution into equation \eref{eq:C03} yields
\begin{equation}
\label{eq:C05}
\fl \sum_{\lambda=1}^{\infty}
\frac{k_{m\lambda}^2}{(k^2-k_{m\lambda}^2)^2} = \frac{R^2}{4
  J_{m}^{2}(nkR)} \left( J_{m}^{2}(nkR) - J_{m+1}(nkR) J_{m-1}(nkR)
\right) \, .
\end{equation}
After substitution into equation \eref{eq:C01} one arrives at the
integrated density of states \eref{eq:ldos04}.

\end{document}